\documentclass[preprint,showpacs,preprintnumbers,superscriptaddress,amsmath,amssymb]{revtex4} 

\usepackage{graphicx}
\usepackage{dcolumn}
\usepackage{bm}
\usepackage{times}

\begin{document}

\title{A stable fiber-based Fabry-Perot cavity}


\author{T. Steinmetz}
\affiliation{Department f\"{u}r Physik,
Ludwig-Maximilians-Universit\"{a}t, Schellingstrasse 4/III,
80799 M\"{u}nchen, Germany}

\author{A. Balocchi}
\affiliation{School of Engineering and Physical Sciences,
Heriot-Watt University, Edinburgh EH14 4AS, UK}

\author{Y. Colombe}
\affiliation{Department f\"{u}r Physik,
Ludwig-Maximilians-Universit\"{a}t, Schellingstrasse 4/III,
80799 M\"{u}nchen, Germany}

\author{D. Hunger}
\affiliation{Department f\"{u}r Physik,
Ludwig-Maximilians-Universit\"{a}t, Schellingstrasse 4/III,
80799 M\"{u}nchen, Germany}

\author{T. W. H\"{a}nsch}
\affiliation{Department f\"{u}r Physik,
Ludwig-Maximilians-Universit\"{a}t, Schellingstrasse 4/III,
80799 M\"{u}nchen, Germany}

\author{R. J.\ Warburton}
\affiliation{School of Engineering and Physical Sciences,
Heriot-Watt University, Edinburgh EH14 4AS, UK}

\author{J. Reichel}
\affiliation{Laboratoire Kastler Brossel de l'E.N.S., 24 rue Lhomond,
  75231 Paris Cedex 05, France}

\date{\today}



\begin{abstract}
  We report the development of a fiber-based, tunable optical cavity
  with open access. The cavity is of the Fabry-Perot type and is
  formed with miniature spherical mirrors positioned on
  the end of single- or multi-mode optical fibers by a transfer
  technique which involves lifting a high-quality mirror from a smooth
  convex substrate, either a ball lens or micro-lens. The cavities
  typically have a finesse of $\sim 1,000$ and a mode volume of
  600\,$\mu$m$^3$.  We demonstrate the detection of small ensembles of
  cold Rb atoms guided through such a cavity on an atom chip. 
\end{abstract}

\maketitle

An optical cavity amplifies the interaction between light and matter
by recirculation of the light at a resonant frequency. This feature is
exploited in a number of fields, notably lasers and optical sensors.
Furthermore, it is crucial to experiments and possible technologies
based on exploiting the quantum mechanical properties of individual
atoms and photons. In this field of cavity quantum electrodynamics
(CQED), the crucial features of the cavity are a small mode waist
and/or mode volume $V_m$, and a high finesse
$\mathcal{F}=\Delta\nu/\delta\nu$ (where $\Delta\nu$ is the free
spectral range and $\delta\nu$ the linewidth of the cavity),
equivalently a high $Q$ factor $Q=\nu/\delta\nu$ \cite{Kimble98}. The
``gold standard'' for CQED cavities is still being set by macroscopic
Fabry-Perot (FP) cavities with superpolished, concave mirrors. These
mirrors have relatively large radii of curvature ($R=20\,$cm is
typical) and achieve record finesse values of $\mathcal{F}> 2\times
10^6$ \cite{Rempe92}. However, there are situations where cavities of
this type cannot be used, and a number of alternative cavity designs
have been proposed (see for example \cite{Spillane05}).  A notable example
calling for an alternative cavity design is single-atom detection for quantum
information processing on atom chips
\cite{Reichel02,FolmanReview02,Horak03,Long03} where the macroscopic cavity 
is incompatible with the microscopic chip. Another example concerns the insertion of quantum objects such as quantum dots and molecules into cavities where the need for cryogenic temperatures makes a macroscopic design problematic. While microscopic cavities are being actively developed, existing designs typically lack an easy way of tuning the cavity into resonance.

\begin{figure}
\begin{center}
\includegraphics[width=100mm]{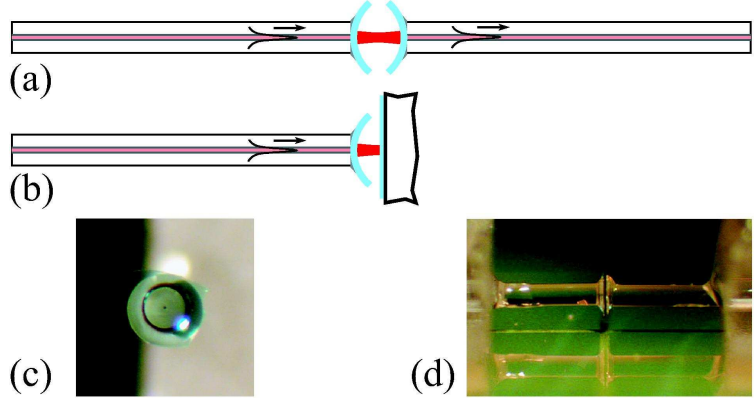}
\end{center}
\caption{\label{fig:concept}(a) and (b) Concept of the miniature
  cavity. The basic building block is an optical fiber functionalized
  with a concave dielectric mirror. Two such fibers, brought
  sufficiently close to each other, result in a stable Fabry-Perot
  cavity which can be interrogated remotely, either in transmission or
  in reflection, through the two fibers (``2FFP'' configuration, (a)).
  Alternatively, a single fiber can be brought close to a reflecting
  planar surface (``1FFP'' configuration, (b)). The 1FFP configuration
  is suitable for use with nanofabricated structures such as quantum
  dots. (c) A single-mode optical fiber, total diameter 125 $\mu$m
  processed with a concave mirror. The mirror has radius 1000 $\mu$m
  with a stopband centered at 780 nm.  (d) A miniature cavity,
  realizing the configuration (a), mounted on an atom chip used in the
  detection of cold atoms (Fig.~\ref{fig:atoms}).}
\end{figure}

An optical fiber-based resonator offers an attractive way forward
\cite{Trupke05,Long03}. Here we describe a fiber-coupled Fabry-Perot
(FFP) cavity that employs concave dielectric mirror coatings with
small radius of curvature, realized on the fiber tip. A stable cavity
is constructed as shown in Fig.~\ref{fig:concept}(a),(b), either from
one such fiber tip facing a planar mirror (``1FFP''), or from two
closely spaced fiber tips placed face-to-face (``2FFP'').  Thus,
unlike the design of \cite{Trupke05}, the cavity can easily be used in
transmission, and does not require a concave depression to be
fabricated on the substrate. We demonstrate how such cavities can be
easily fabricated using commercially available lift-off coatings.
Tunability is achieved by attaching one of the fibers to a
piezoelectric actuator.  Because of the small fiber diameter
(125\,$\mu$m), very short cavities ($<10 \lambda/2$) can be realized
even with radii of curvature $R\le1\,$mm, still leaving a sufficiently
large gap to introduce cold atoms. We achieve stable cavity modes with
a finesse of $\sim 1,000$ in the near infrared. The mode volume is
$V_m=600\,\mu\textrm{m}^3$, to be compared to
$V_m=1680\,\mu\textrm{m}^3$ for the smallest-volume macroscopic FP
cavity that has been used with atoms \cite{Hood00}.  In terms of CQED
parameters, the small mode volume results in an exceptionally high
coherent atom-photon coupling rate, $g_0/2\pi=180\,$MHz (calculated
for the Rb D2 line at $\lambda=780\,$nm) . Therefore, in spite of the
comparatively high damping rate, $\kappa/2\pi=2.65\,$GHz which results
from the moderate finesse and short length, the cavity reaches a
single-atom cooperativity parameter greater than unity,
$C=g_0^2/2\kappa\gamma=2.1$ where $\gamma/2\pi=3\,$MHz is the atomic
linewidth, signaling the onset of quantum effects such as enhanced
spontaneous emission into the cavity mode and a significant
modification of cavity transmission by the presence of a single atom.
The potential of this approach is demonstrated here with an experiment
using a 2FFP cavity to detect an extremely weak flux of cold atoms
magnetically guided on an atom chip.  We present this technology not
only as an important stepping stone towards on-chip single atom
detection but also for cavity experiments with quantum dots,
semiconductor nanocrystals and molecules, and for telecoms devices.

The concave mirrors are fabricated from a convex template and a
lift-off step. For large radii of curvature, $\ge 500$ $\mu$m, the
template is a commercial ball lens whereas for smaller radii,
$100-500$ $\mu$m, the template is a silica micro-lens specially
fabricated for these experiments. The micro-lenses are etched into a
planar silica substrate following the melting and re-solidification of
a photoresist mesa. Surface tension provides an extremely smooth
photoresist surface such that the roughness in the micro-lens is
determined only by the subsequent dry-etching step and can be as small
as $\sim 1$ nm. The template is
coated \cite{OIB} in one run with a
release layer and silica-titania dielectric Bragg stack, with a
stopband centered either at 780 nm or 850 nm and nominal reflectivity of 99.7\%. 
We then position a
cleaved single mode fiber immediately above the center of the coated
lens by maximizing the back reflection of a laser beam coupled into
the fiber. The fiber is then glued in place with an UV-curing epoxy,
after which the application of a small force is sufficient to detach
the mirror from the original substrate. The result is a fiber
functionalized with a highly reflecting concave mirror, as shown in
Fig.~\ref{fig:concept}(c). A complete 2FFP cavity is shown in
Fig.~\ref{fig:concept}(d).

\begin{figure}
\begin{center}
\includegraphics[width=100mm]{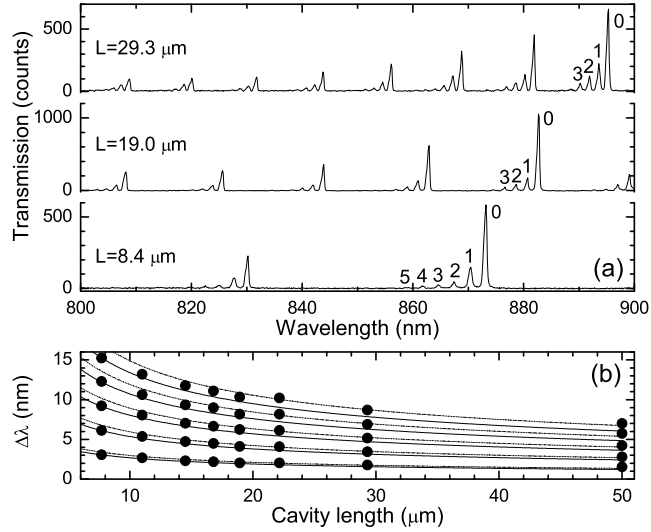}
\end{center}
\caption{\label{fig:white}(a) White light transmission spectra of a
  1FFP cavity recorded at three different cavity lengths, $L$. The mirrors
  have a stopband centered at 850 nm. $L$ is the effective cavity
  length, determined by $L=1/2 \Delta(1/\lambda)$ where
  $\Delta(1/\lambda)$ is the change in wave number from one
  fundamental longitudinal mode to the next. The modes are labeled
  with the sum of the two lateral mode indices, $m$ and $n$. The
  widths of the transmission peaks are limited by the spectrometer and
  therefore do not reflect the true finesse. (b) Separation in
  wavelength of the higher lateral modes from the fundamental mode as
  a function of $L$ at $\lambda=850$ nm. The solid (dashed)
  line represents the analytical result for a spherical-planar
  FP cavity with radius $R=230$ (200) $\mu$m.}
\end{figure}

In order to characterize the modes of a 1FFP cavity, we measured the
white light transmission spectrum for several values of the cavity
length $L$. We find that stable cavity modes with high $\mathcal F$
can be established with little attention to the alignment, unlike the
planar-planar cavity geometry which is extremely sensitive to
alignment and mechanical noise. The results are shown in
Fig.~\ref{fig:white}(a). As expected for a FP cavity, for each $L$
there is a series of longitudinal modes, until at the smallest $L$
there is just one longitudinal mode in the stopband. Each longitudinal
mode consists of a series of finely spaced modes, corresponding to the
different transverse modes. This lifting of the degeneracy of the
transverse modes allows a spectroscopic determination of the mirror
curvature.  Fig.~\ref{fig:white}(b) plots the wavelength shift of the
higher order modes relative to the fundamental as a function of $L$
showing how the spacing increases as $L$ decreases. The solid curves
in Fig.~\ref{fig:white} are the results from Gaussian optics for a
stable, planar-spherical cavity \cite{Yariv89}, $\Delta \lambda =
(\lambda^2/2 \pi L) \Delta(m+n)\cos^{-1} \sqrt{1-L/R}$ where $m$, $n$ are the lateral
mode indices. At large (small) $L$, the results are best described
with a radius of about 200 (230) $\mu$m.  This compares well with the
radius of the micro-lens template, 250 $\mu$m. The slight dependence
on $L$ is likely to result from a slight ``softening'' of the mirror
away from its center: as $L$ increases, the beam waist at the curved
mirror increases, probing more of the curved mirror.

\begin{figure}
\begin{center}
\includegraphics[width=100mm]{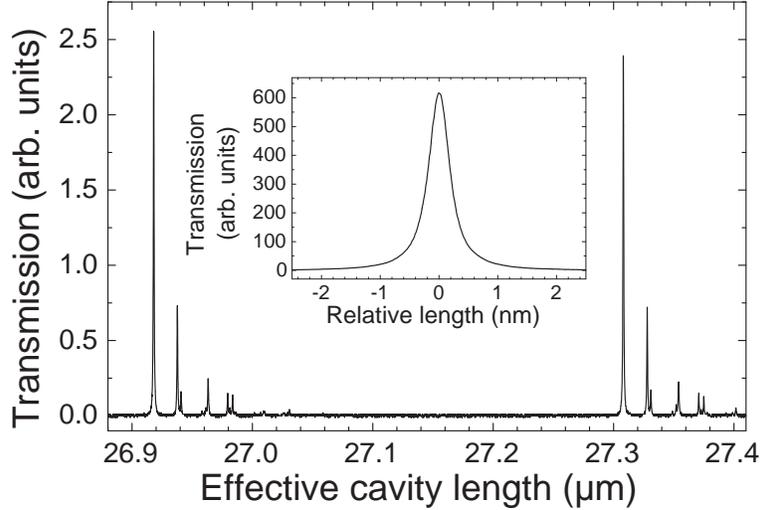}
\end{center}
\caption{\label{fig:FSR}Transmission of a 2FFP cavity versus cavity
  length. A piezo is used to vary the cavity length. The absolute
  length is determined by using up to three lasers as described in the
  text. The mirrors have radius 1,000 $\mu$m with a stopband centered
  at 780 nm. The cavity has an effective cavity length
  of 27\,$\mu$m with a finesse of 1050; equivalently, free spectral
  range 5500 GHz and mode width (FWHM) 5.29 GHz. The inset shows the
  line shape of the fundamental (0,0) mode.}
\end{figure}

The limited resolution of the spectrometer prohibits a proper
measurement of the finesse, and only a lower limit can be deduced
($\mathcal F>500$ for small $L$). We have therefore measured the
cavity transmission as a function of $L$ using grating-stabilized
diode lasers (Fig.~\ref{fig:FSR}). To determine the finesse, two
lasers were simultaneously coupled into the cavity.  The first laser
is locked to a sub-Doppler line in the $^{87}$Rb D2 spectrum at
$\lambda=780.27$\,nm, the second is tuned several $\kappa$ away to
$780.6\,$nm using a wavemeter. The known wavelength difference of the
two lasers allows us to calibrate the length scan with a better
precision than the length-voltage characteristics of the piezo.
Additionally, we simultaneously recorded the transmission of a third
diode laser at 828.25\,nm in order to determine unambiguously the
absolute cavity length. As a typical result, we have measured a
finesse $\mathcal F=1,050$ for a 2FFP cavity with $L=27\,\mu$m and
$R=1,000\,\mu$m using a single-mode fiber on the input side and a
multimode fiber on the output side.  This is in good agreement with
independent measurements of the mirror transmission, $T=8\times
10^{-4}$, and loss, $\ell=2.4\times 10^{-3}$ \cite{loss}. From these
values, the expected finesse is $\mathcal F=\pi/(T+\ell)=980\pm 40$.
This indicates that the finesse of the FFP is as high as the coatings
allow.  The 1/e beam waist in the cavity in Fig.~\ref{fig:FSR} is
$w=5.4\,\mu$m, implying a cavity mode volume $V_m\sim 600$ $\mu$m$^3$
and $C=2.1$ for Rb.

\begin{figure}
\begin{center}
\includegraphics[width=100mm]{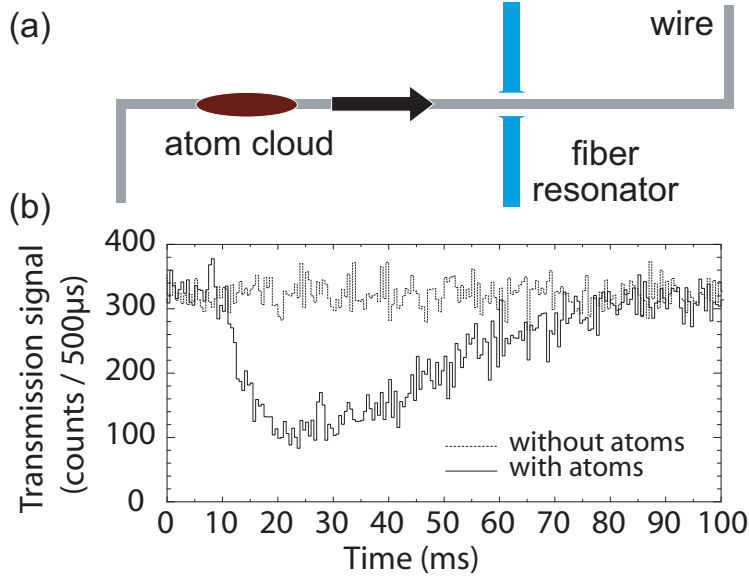}
\end{center}
\caption{\label{fig:atoms}Atom detection with an on-chip fiber
  resonator.  (a) At $t=0$, a magnetically trapped atom cloud
  ($T=70\,\mu$K) is released into a very elongated Ioffe-Pritchard
  potential, created using the wire shown in gray. This potential
  guides the atoms through the center of the resonator mode.  (b)
  Transmission signal of the fiber resonator for a single experimental
  run (solid line), along with an empty cavity transmission signal
  (dashed line).  The transmission drops to 35\% of the
  empty-resonator value.}
\end{figure}

We have used the 2FFP cavity as a very sensitive detector for
magnetically guided atoms on an atom chip. The experimental setup is
similar to our previous experiments \cite{Haensel01a,Du04}, but
contains an FFP subassembly. Each fiber is glued onto a piezoelectric
actuator, and the piezos are glued onto a ceramic bridge while
monitoring the cavity transmission signal. The mirror spacing on axis
is 27\,$\mu$m as in Fig.~\ref{fig:FSR}, leading to a gap between
mirror edges of $\sim 15\,\mu$m; the finesse is 260. The subassembly
is glued onto the chip with a 230\,$\mu$m separation between the
cavity axis and the chip surface (Fig.~\ref{fig:atoms}~(a)). A
magnetically trapped cloud of $^{87}$Rb atoms at a temperature of
70\,$\mu$K is produced in an initial trap and then released into a
very elongated Ioffe-Pritchard potential, created using a ``Z
wire'' \cite{Reichel02}.  This potential guides the atoms through the
center of the resonator mode. The cavity mode is excited by a very
weak resonant probe laser, both the laser and the mode being tuned to
atomic resonance.  The transmitted signal is detected with a photon
counter.  Fig.~\ref{fig:atoms}~(b) shows sample transmission signals
with and without atoms. We have independently determined the atomic
density and temperature before entry into the cavity using absorption
imaging and time-of-flight analysis. Integrating the initial density
over the cavity mode yields an upper-bound estimate $N_{\max}$ for the
number of atoms in the cavity mode, $N_{\max}\sim 50$.  It is clear
however that the true number of atoms contributing to the signal is
much lower because the transverse size of the atom cloud is
significantly larger than the gap between the mirrors, so that a large
part of the atoms is lost upon entering the cavity, and does not
contribute to the signal.

Our current cavities aim for high cooperativity. It is interesting however to consider what improvements would be necessary to
enter the strong coupling regime of CQED, i.e., $g_0>\kappa,\gamma$
\cite{Kimble98}.  To
obtain strong coupling, it is preferable to increase the mirror
distance $L$ provided $L\ll R$: for a given $R$, $\kappa$ drops as
$\kappa\propto L^{-1}$, whereas $g_0$ only decreases as $g_0\propto
L^{-3/4}$.  $L=200\,\mu$m is a realistic value where alignment should
still be manageable, and where the spot size on the mirrors remains
much smaller than the mirror diameter, so that clipping loss can still
be neglected. At this $L$, the parameters of the 2FFP cavity with
$R=1\,$mm would become $g_0/2\pi=42\,$MHz and $\kappa/2\pi=360\,$MHz.
Thus, the finesse needs to be improved by roughly a factor of 10, to
$\mathcal F=10,000$, in order to reach the strong coupling regime.
Presently, the finesse of our cavities is limited by the quality of
the multilayer coatings. However, the transfer coating technology
continues to improve as suggested by the fact that the transfer
coatings used in \cite{Trupke05}, fabricated after the ones used here,
resulted in a measured finesse $\mathcal F=6,000$.  Speculating that
the coating quality can be improved further, the remaining issue is
the surface roughness of the template for the liftoff process. For a
ball lens from the batch used in our current cavities, an AFM
measurement gave an rms roughness of $\sigma=1.7\,$nm. Following a
standard estimate for scattering loss, $S=(4\pi\sigma/\lambda)^2$,
this roughness must be improved to $\sigma\le 0.7\,$nm which is
achievable both with superpolishing and with micro-lens fabrication.

We conclude by noting that, even without any further improvement, the
FFP cavities described here should enable single-atom detectivity with
good signal-to-noise ratio \cite{Horak03,Long03}. Unlike
dispersive detection without a resonator \cite{Lye03}, this technique
will ultimately allow detection with less than one spontaneous
emission per atom on average, enabling preparation of single-atom
states. 

A.B. and R.J.W. acknowledge financial support from EPSRC (UK) and
the Royal Society (London). The remaining authors acknowledge support
from the EU (project IST-2001-38863, ACQP) and from the Bavarian State
Government (\emph{Kompetenznetzwerk Quanteninformation -- A8}). Y.C.
thanks the EU CONQUEST network (MRTN-CT-2003-505089) for his stipend.


\begin{thebibliography}{13}
\bibitem{loss}
The loss measurement was performed on coatings produced in the same run, 
but on two macroscopic high-quality concave substrates, by measuring the 
finesse of a FP cavity constructed with these mirrors.
\end{thebibliography}

\end{document}